\documentclass[sn-mathphys-num]{sn-jnl}

\usepackage[utf8]{inputenc}
\usepackage[T1]{fontenc}

\usepackage{graphicx}%
\usepackage{multirow}%
\usepackage{amsmath,amssymb,amsfonts}%
\usepackage{amsthm}%
\usepackage{mathrsfs}%
\usepackage[title]{appendix}%
\usepackage{xcolor}%
\usepackage{textcomp}%
\usepackage{manyfoot}%
\usepackage{booktabs}%
\usepackage{algorithm}%
\usepackage{algorithmicx}%
\usepackage{algpseudocode}%
\usepackage{listings}%

\usepackage{amsmath, amsthm, amscd, amsfonts, amssymb, graphicx, color,colortbl, latexsym}
\usepackage{textcomp}
\usepackage{graphicx,subfigure}
\usepackage{array}
\usepackage{caption}

\usepackage{booktabs}

\usepackage{algorithm}
\usepackage{algpseudocode}
\usepackage[utf8]{inputenc}
\usepackage[english]{babel}
\usepackage{extpfeil}
\usepackage{tabularx}
\usepackage{balance}
\usepackage{lineno,hyperref}
\usepackage{lipsum}
\usepackage{soul}

\theoremstyle{thmstyleone}%
\theoremstyle{thmstyletwo}%

\theoremstyle{thmstylethree}%

\begin{document}
	
	\title[A Blockchain-Based Quality Control Model for Online Collaboration Systems]{A Blockchain-Based Quality Control Model for Online Collaboration Systems}
	
	
	\author[1]{\fnm{Sadegh} \sur{Sohani}}\email{sd\_sohani@mail.um.ac.ir }
	
	\author[1]{\fnm{Maliheh} \sur{Shahryari}}\email{maliheh.shahryari@mail.um.ac.ir}
	
	\author[2]{\fnm{Salar} \sur{Ghazi}}\email{seyed-salar.ghazi.1@ens.etsmtl.ca}
	\author*[1]{\fnm{Mohammad} \sur{Allahbakhsh}}\email{allahbakhsh@um.ac.ir}
	\author[1]{\fnm{Haleh} \sur{Amintoosi}}\email{amintoosi@um.ac.ir}
	\author[3]{\fnm{Boualem} \sur{Benatallah}}\email{boualem.benatallah@dcu.ie}
	\affil*[1]{\orgdiv{Computer Engineering Department}, \orgname{Ferdowsi University of Mashhad}, \orgaddress{\city{Mashhad}, \country{Iran}}}
	
	\affil[2]{\orgname{\'Ecole de technologie sup\'erieure}, \orgaddress{ \city{Montr\'eal}, \state{Qu\'ebec}, \country{Canada}}}
	
	\affil[3]{\orgdiv{School of Computing}, \orgname{Dublin City University}, \orgaddress{ \city{Dublin}, \country{Ireland}}}
	
	\abstract{Collaborative content generation (CCG) enables collective creation of artifacts like scientific articles. Quality is a paramount concern in CCG, and a multitude of methods have been proposed to evaluate the quality of artifacts. Nevertheless, the majority of these methods are reliant on centralized architectures, which present challenges pertaining to security, privacy, and availability. Blockchain technology proffers a potential resolution to these challenges, by furnishing a decentralized and immutable ledger of quality scores. In this manuscript, we introduce a blockchain-based quality control model for CCG that uses a semi-iterative algorithm to interdependently compute quality scores of artifacts and reputation of nodes. Our model addresses critical challenges in academic informetrics, such as citation manipulation, transparency in collaborative scholarship, and decentralized trust in metric computation. Our model also exhibits sensitivity to processing latency, rendering it more agile in the presence of delays. Our model’s quality scores, evaluated against PageRank and HITS baselines, show comparable performance, with additional assessments of throughput, latency, and robustness against malicious nodes confirming its reliability. A theoretical comparison with recent studies validates its feasibility for real-world informetric application.}
	
	
	
	
	\maketitle
	
	\section{Introduction}
	Collaborative problem-solving and decision-making is known as ''Collective Intelligence (CI)''\cite{1}. This concept has been applied to various research domains, such as sociology, psychology, biology, tourism, management, economics, computer science, game theory, and more~\cite{2,3,4,5}. The development of information and communication technologies, such as Web 2.0, semantic web, and crowdsourcing, has enabled large-scale collaborations in the online space. These technologies allow individual users and groups to share knowledge, solve problems, and make decisions through web-based interactions and collaborations. In a specific form of such collaborations, collaborative content generation (CCG), a group of contributors work together to create content~\cite{1,5,6}. We refer to such collaboratively created content as artifacts. Some examples of CCG are scientific articles, Wikipedia, and Github.
	
	In CCG systems, there are two forms of collaborations: synchronous and asynchronous~\cite{7}. In synchronous collaboration, contributors collaborate intentionally and simultaneously in the content creation process. Writing research articles is an example of synchronous collaboration, where researchers work together on the same document. In asynchronous collaboration, an artifact is created by one contributor, and then modified by other contributors over time, without explicit coordination. These artifacts can be original or derived from existing ones, creating parent-child relationships among them. Wikipedia and Github are examples of asynchronous collaborative systems~\cite{7,8,9}.
	
	Blockchain technology provides a decentralized, tamper-proof framework for managing data through a distributed ledger, where transactions are recorded immutably across a network of nodes. Various ledger structures, such as linear blockchains and Directed Acyclic Graphs (DAGs), enable flexible data management tailored to specific application needs. Consensus algorithms ensure agreement among nodes on the ledger’s state, enabling trustless collaboration without a central authority~\cite{bcsurvey,sciot2024cons}. 
	
	One of the main challenges of CCG systems is how to ensure the quality of the content~\cite{10,11}. Quality control is difficult in CCG systems, as the contributors may have different backgrounds, expertise, motivations, and incentives. Moreover, the content may be complex, dynamic, and interrelated, requiring constant evaluation and revision. The proliferation of preprint repositories, open peer review platforms, and collaborative authorship tools underscores the need for decentralized, tamper-proof quality assessment, a gap our blockchain-based model seeks to fill. 
	Various techniques and methods have been proposed to address this challenge, by leveraging human feedback, content, the network structure and the relationships between entities to calculate dependable quality metrics. The earlier approaches, simple aggregations, weighted aggregations, and heuristics attempted to aggregate direct or indirect human feedback to compute quality scores~\cite{QQ,AQA,wikitrust}. However, these techniques have several limitations. Mainly, they are prone to manipulation.
	
	To overcome this problem, iterative techniques are proposed, in which network structure and the relationships between entities are employed to calculate interdependent quality metrics, making them more resistant to manipulation than simple or weighted aggregations. Some examples of such systems are PageRank~\cite{pagerank}, RTV~\cite{RTV1}, Dekerchove~\cite{Dekerchove}, Laureti~\cite{laureti}, and SciMet~\cite{7}. However, these systems are mostly designed with centralized architectures, which result in problems such as single point of failure, lack of transparency, and high computational overhead.
	
	Systems based on blockchain technology present a promising solution to address these challenges. These systems are devoid of a singular central node, implying that even in the event of failure of one or multiple computational nodes or servers, the remaining nodes retain the ability to compute metrics and disseminate scores. Furthermore, blockchain technology facilitates collaboration among individuals lacking mutual trust, eliminating the necessity for a central and impartial authority~\cite{bcsurvey}. Owing to the distributed architecture of blockchain, any modifications to the data necessitate network consensus, thereby bolstering security against attacks and attempts at data alteration. These attributes render blockchain a unique technology with particular relevance in CCG systems.
	
	Nevertheless, systems based on blockchain technology are not without their challenges. Primarily, iterative techniques are computationally intensive and time-consuming, making their execution on blockchain nodes challenging. Managing a consensus on their results also proves to be a time-consuming task. In addition, the consensus techniques currently employed in scoring systems do not adapt well to processing latency. As delay escalates, these techniques encounter difficulties and fail to reach a consensus.
	
	In addition to these challenges, CCG systems face a major obstacle concerning content quality control on a public blockchain. Sharing raw content in a completely decentralized environment can be problematic, as untrusted nodes would gain access to sensitive or confidential information. Consequently, previous approaches have often relied on private blockchains or centralized methods to ensure data confidentiality. In contrast, our model leverages a public blockchain by employing a PageRank-like mechanism that uses citation relationships among artifacts, thus avoiding the direct sharing of full content. Each artifact’s score is calculated using this citation-based approach, combined with a specialized consensus protocol. This protocol applies a genetic algorithm to select a committee of nodes, ensuring that the evaluation process remains robust, decentralized, and resistant to manipulation while still maintaining the privacy of raw data.
	
	This manuscript presents our model for quality control in CCG systems, designed to address the aforementioned issues. We propose an innovative data model that encapsulates and simplifies the comprehension of the research problem and our suggested solution. The data model is composed of a directed acyclic graph, which represents the contents and their interrelationships, and a blockchain that functions as the consensus infrastructure.
	
	We formulate an adaptive semi-iterative algorithm that operates on each node independently, calculating scores. This algorithm incorporates a convergence threshold that dictates the termination of the algorithm. This threshold is adapted to the process latency, while ensuring the system accuracy remains above a predefined acceptable threshold.
	
	Subsequently, the computed scores are dispatched to a committee of selected nodes as proposals, to be aggregated and reach a consensus. We propose a committee-based consensus algorithm that aggregates the scores received from committee nodes and generates a quality score for each content. These scores are then disseminated through the network and recorded on the local ledgers. To summarize, the main contributions of the paper are as follows:
	\begin{itemize}
		\item
		We introduce a meta model that simplifies the comprehension of the problem and the proposed solution. The model comprises a directed acyclic graph and a blockchain.
		\item
		We formulate a semi-iterative method for computing quality scores on each node of the blockchain.
		\item
		We put forth an innovative model for managing the propagation of updates within the artifact graph, designed to adapt to processing latency while maintaining an acceptable level of accuracy.
		\item
		We present a novel committee-based consensus algorithm that aggregates scores received from committee members.
		\item
		We evaluate the model’s performance and demonstrate that its results are consistent and meaningful when compared to two well-known baseline models (PageRank and HITS)
	\end{itemize}
	The rest of the paper is structured as follows: In Section~\ref{sec:rels}, we review the related literature. We introduce our data model and an example scenario in Section~\ref{sec:basic}. In Section~\ref{sec:model}, we present our proposed framework. We evaluate the performance of our framework in Section~\ref{sec:eval}, and finally, we conclude in Section~\ref{sec:conc}.
	
	\section{Related Work}\label{sec:rels}
	Quality assessment constitutes a significant challenge within the realm of collaborative systems. This section elucidates this challenge and provides an overview of the research endeavors in this domain. The pertinent works are reviewed from the perspectives of quality measurement studies in both centralized and decentralized systems.
	
	The authors in~\cite{9} undertake an exhaustive analysis of quality control in the process of eliciting information from a substantial population. Specifically, they probe into the methodologies to guarantee quality in the process of gathering information from individuals. The study delves into the constituents of quality, techniques, and measures to assure quality as a consequence of this investigation. The paper furnishes details such as quality attributes that are indispensable in the information elicitation process, quality assessment techniques employed to gauge the degree of quality, and measures implemented to ensure quality in this process.
	
	In~\cite{12}, the authors introduce a framework termed the ''Comprehensive Researcher Achievement Model (CRAM)'' designed to evaluate the accomplishments, impact, and influence of distinguished researchers. This framework is envisaged as an instrument for assessing the impact and efficacy of researchers. It takes into account a variety of information types, encompassing quantitative measurements such as H-index indicators, as well as qualitative factors like involvement in research projects. In~\cite{27}, a comparative analysis is conducted among the top 10 authors who have incorporated celiac disease as a keyword in Google Scholar, utilizing three authorship indices: h-index, c-index, and c'-index. The h-index ranks authors predicated on the number of publications (h) that have received h or more citations. In the computation of the c-index, the author's position influences the weight of authorship contribution. In the c'-index, both the author's position and the total number of authors are factored into the weighting of the author's contribution.
	
	The authors in~\cite{28} undertake a study to evaluate the correlation between citations, impact, and quality, illustrating that the number of citations can be detrimental to numerous articles due to the perception of inferior quality. That is, equating the number of citations to quality can result in an overestimation of the quality of highly cited articles. ~\cite{pagerank} introduces the renowned concept of ranking web pages in the PageRank algorithm, where the quality and quantity of pages linked to a page are reconsidered for ranking the page. It is posited that the PageRank is resistant to manipulations due to the consideration of global variables that are more robust against potent manipulations. PageRank has inspired a multitude of ranking methods known as schemes inspired by PageRank, such as Eigenfactor and SCImago Journal Rank (SJR)~\cite{eigenfactor}.
	
	The authors in~\cite{7} put forth an iterative definition wherein the quality of artifacts, contributors, and venues are defined in an interdependent manner. Within this framework, the quality of an artifact is determined based on the quality of its contributors, venue, references, and citations. The quality of a contributor is ascertained by the quality of their artifacts, contributors, and venues. The quality of a venue is defined predicated on the quality of its artifacts and contributors. In~\cite{31}, an iterative model for quality in a synchronous collaborative content generation system is introduced. This model takes into account multiple factors such as popularity, community attention, and relationships between artifacts and contributors. 
	
	The authors in~\cite{32} conduct an exploration of blockchain-based social media platforms with financial incentives, utilizing the Steemit platform as a case study. These platforms employ blockchain technology to incentivize users to create and engage with content, detailing the distribution of financial incentives to users. The study also investigates the positive and negative effects of this incentive system on content production and online communities. Austin et al.~\cite{new4} and Alshehri et al.~\cite{new5} apply blockchain to domains of decentralized finance and e-voting, respectively, highlighting the broad applicability of blockchain for trust and scoring mechanisms.
	
	The authors in~\cite{33} conduct an investigation into a blockchain-based evaluation system for the validation of the safety and efficacy of healthcare treatments. This system leverages blockchain technology to ensure transparency, security, and trust in the assessment process of healthcare treatments. The authors in~\cite{21} explore a blockchain framework termed ''ExCrowd'' for information gathering via an Exploration-Based approach. This is a crowdsourcing system that employs a smart contract as a reliable reference for the appropriate selection of individuals, evaluation of inputs, and provision of incentives while preserving user privacy. In~\cite{34}, a proposal is made for a blockchain-based content ranking system for an online review and ranking system in the context of online education. This system suggests a decentralized and trustworthy platform that guarantees the integrity of content ranking and the independence and integrity of content reviews by subject matter experts. In~\cite{new6}, the authors tackle NFT fraud in blockchain traceability systems by proposing a credit evaluation method using the Analytic Hierarchy Process (AHP) to assess user authenticity before data is recorded, addressing the untraceable pre-blockchain data gap. By analyzing multiple indexes to create an evaluation matrix and efficacy coefficient, the system effectively rates user credit, enhancing trust and security in NFT markets, akin to~\cite{34}’s focus on reliable blockchain-based ranking systems. 
	
	In~\cite{new7}, the authors propose a blockchain-based university ranking system using Ethereum smart contracts to enhance transparency and trust, overcoming the subjectivity and centralization issues of traditional systems like QS and ARWU. In~\cite{new9}, the authors introduce AssessChain, a hybrid blockchain system combining a consortium and public blockchain to enhance transparency and reliability in online assessments, using smart contracts and IPFS for secure, efficient execution and storage. The authors in~\cite{new10} present a permissioned blockchain framework on Hyperledger Sawtooth to rank synthetic data generators, addressing selection suitability and transparency by integrating a purpose-driven algorithm that evaluates utility and privacy risks. In~\cite{new3}, the authors  conduct a systematic meta-analysis using the PRISMA-P methodology to evaluate blockchain’s applications in Education 4.0, emphasizing its role in enhancing educational systems while identifying security gaps beyond confidentiality, integrity, and availability. 
	
	The authors in~\cite{35} propose a blockchain-based Cyber Threat Intelligence (CTI) information system architecture that collects, assesses, stores, and shares CTI. It facilitates the possibility of tamper-resistant and deletion-proof assessments of untrusted CTI feeds while evaluating the quality of CTI feeds against a defined set of quality standards. The data evaluation is conducted using a reputation-based mechanism to select assessors, which rank the CTI feeds based on quality parameters. To ensure fair results and their final storage, they introduce a Proof of Quality (PoQ) consensus algorithm. Similarly, in~\cite{ghazi2022suitability}, the authors enhance TrustChain to secure it against whitewashing and client vulnerability issues. In Trustchain, which is a DAG-based blockchain, each agent publishes its unique chain, monitors the interactions of other nodes, and collects TrustChain data to calculate trust levels. In~\cite{new2}, the authors introduce a decentralized CTI sharing platform using Distributed Ledger Technology to counter sophisticated cyber threats like zero-day attacks, ensuring secure, tamper-proof, and transparent data exchange via blockchain’s immutable nature and consensus mechanisms. This approach eliminates centralized trust dependencies, enhances security, reduces sharing latency, and strengthens resilience, addressing limitations of traditional models like data integrity risks and single points of failure. Complementing this, in~\cite{new11}, Venčkauskas et al. propose a Hyperledger Fabric-based model for incentivized CTI sharing, integrating monetary rewards via subscription fees and reputation scores to motivate participation and ensure trust, using IPFS for off-chain storage and smart contracts to manage user activities with reward scores validated by MATLAB simulations, aligning with and enhancing efforts like~\cite{35} for secure, transparent global cybersecurity collaboration.
	
	In summary, existing quality assessment techniques for CCG systems are predominantly centralized, relying on databases prone to manipulation, single points of failure, and transparency deficits, as seen in traditional bibliometric indicators like citation counts and journal impact factors~\cite{7, RTV1, 31, pagerank, Dekerchove}. Decentralized approaches, while leveraging blockchain for transparency and trust in applications like content ranking and peer review~\cite{21, 32, 35, new2, new7, new9, new10}, often face computational constraints that limit iterative scoring methods and lack adaptability to processing latency~\cite{7, 31, RTV1, new11}. Moreover, they rarely address the interplay of direct and indirect endorsements critical for rational quality metrics~\cite{pagerank, eigenfactor}. Also, recent blockchain-based frameworks are often private or use blockchain solely as a rating infrastructure, and there is no public blockchain model to algorithmically calculate the quality of artifacts~\cite{34,35}. Our work tackles these challenges by introducing a blockchain-based, semi-iterative model that enhances transparency, resists manipulation, and adapts to latency, advancing informetric methodologies for robust, decentralized quality assessment in CCG systems.
	
	\section{Principles and Basic Concepts}\label{sec:basic}
	In this section, we present the key concepts used in our proposed model to facilitate the comprehension of our model. We first introduce the data model that captures and simplifies the research problem and our proposed solution. Then, we present the example scenarios that illustrate the application and functionality of our model.
	\begin{figure}[!t]
		\centering
		\includegraphics[scale=0.50]{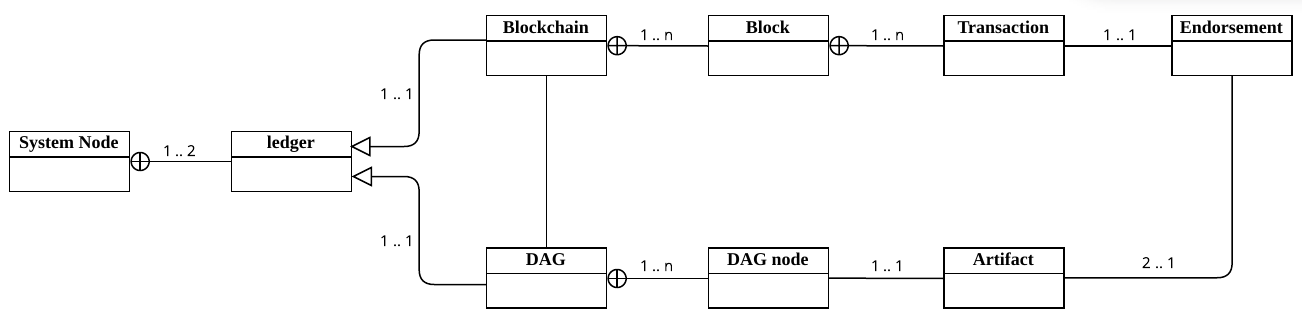}
		\caption{The meta-model of the entities and their relationships}
		\label{fig:mm}
	\end{figure}
	
	\subsection{Meta Model}\label{ssec:dm}
	Consider a quality control mechanism for CCG systems, embodied as a blockchain-based system, comprising a collection of nodes that participate in the quality control process. In a CCG system, content generated by contributors, are represented by artifacts. Denoted as $A=\{a_i\}$, each artifact $a_i$ has a unique identifier and an associated quality score $S_i$. The initial quality score, $S_i^{(0)}$, remains consistent for all new artifacts.
	
	An endorsement link represents a directed edge connecting two artifacts in the \emph{Directed Acyclic Graph (DAG)}. When artifact $A$ endorses artifact $B$ (e.g., through citations or links), it implies a positive relationship. Denoted as $i \rightarrow j$, this link signifies that artifact $a_i$ endorses $a_j$. Each endorsement link has a unique identifier and a weight reflecting its strength. The number of outgoing endorsement links from an artifact $a_i$, is denoted as $L_i$.
	
	Figure~\ref{fig:mm} illustrates the meta model of the proposed Collaborative Content Generation (CCG) system. In this model, the entities involved in the quality control process are denoted as \emph{system nodes}. Each system node maintains two distinct local \emph{ledgers}: a conventional \emph{blockchain} ledger and a local DAG ledger. The blockchain ledger, structured as a chain of \emph{blocks}, records endorsements and ensures data immutability. Each block within the blockchain comprises a series of \emph{transactions}, with each transaction encapsulating an \emph{endorsement} relationship between two artifacts.
	
	A Directed Acyclic Graph in blockchain is a data structure that enables faster consensus and transaction retrieving. In our model the DAG ledger captures the dynamic state of the system, containing one or more \emph{DAG nodes}. Each DAG node represents an individual \emph{artifact}, facilitating the representation of the most recent state of the corresponding blockchain. The DAG reflects bibliographic pair networks, where artifacts represent scholarly works and edges encode citation relationships.  This dual-ledger approach enables the system to efficiently manage and verify the quality of contributions within the CCG framework.
	
	\subsection{Application Scenarios}\label{ssec:scenario}
	Our proposed model is designed to enhance transparency and reliability in assessing artifact quality within CCG systems, with significant implications for informetric research. By leveraging citation relationships and decentralized consensus, the model addresses key challenges in scholarly metrics, such as manipulation and lack of trust in centralized systems. Below, we illustrate its applicability through two informetric scenarios: ranking scientific articles and evaluating open peer review contributions.
	
	\subsubsection{Ranking Scientific Articles}
	In scientometric analysis, ranking scientific articles based on citation networks is a cornerstone of impact assessment. Our model treats articles as artifacts within a DAG, where citation links represent endorsements that contribute to quality scores. When article $A$ cites article $B$, our semi-iterative algorithm (Section~\ref{sec:preparing-proposals}) computes $B$’s quality score by aggregating weighted contributions from $A$’s score, adjusted by a damping factor and propagated through the DAG. This decentralized approach, validated against PageRank and HITS (Section~\ref{sec:experimental-eval}), mitigates citation manipulation by requiring consensus across blockchain nodes, ensuring tamper-proof scores. While other parameters exist for ranking scientific articles~\cite{7}, we highlight the citation count model due to its popularity and relevance to our approach~\cite{AQA}.
	
	\subsubsection{Evaluating Open Peer Review Contributions}
	Open peer review platforms, such as $F1000Research$ and $Publons$, foster collaborative evaluation of scholarly artifacts, but assessing the quality of reviewer contributions remains challenging. Our model represents reviews as artifacts, with endorsements derived from author responses, community ratings, or subsequent citations of the reviewed article. Quality scores for reviews are computed based on these relationships, using our blockchain-based consensus mechanism to ensure transparency and resistance to bias. For instance, a review endorsed by multiple stakeholders receives a higher score, reflecting its scholarly value. By recording scores on a distributed ledger, our model provides a verifiable record of reviewer impact, enabling informetric analysis of peer review contributions.
	
	\section{Proposed Model}\label{sec:model}
	
	\begin{figure}[!t]
		\centering
		\includegraphics[scale=0.35]{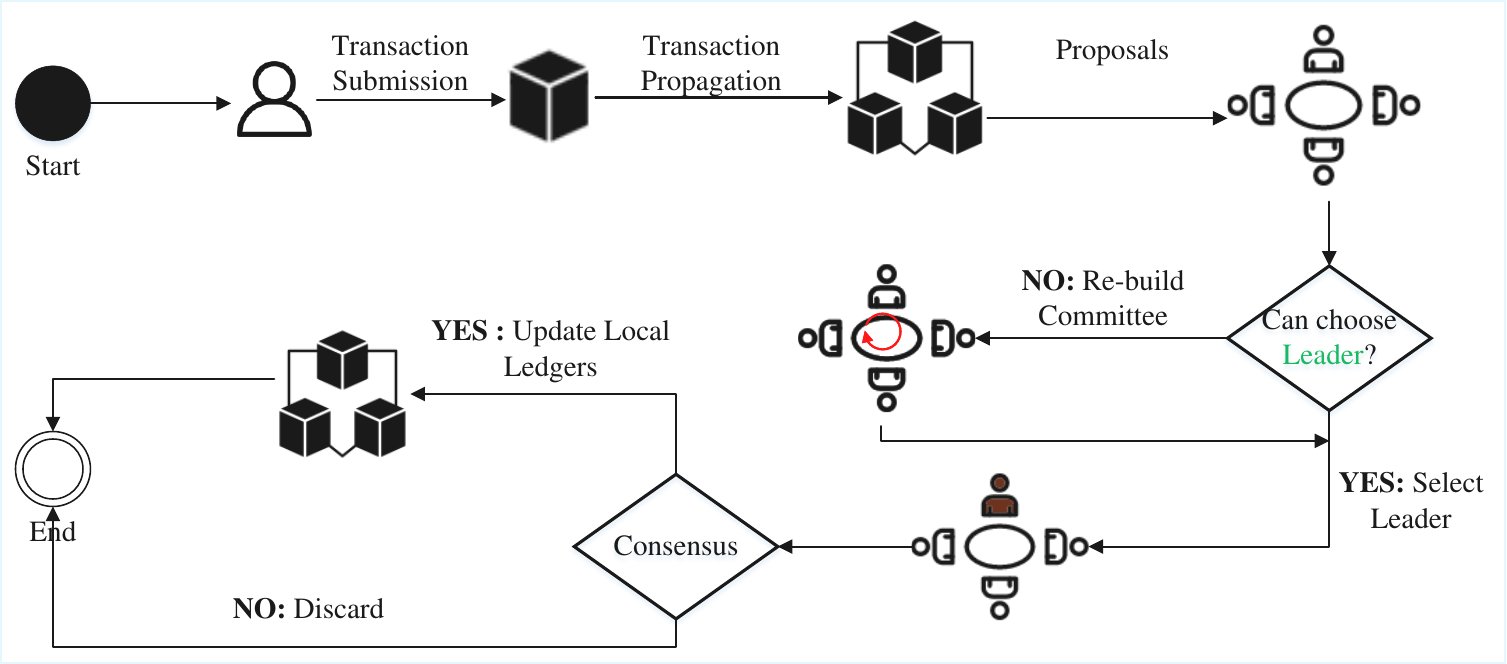}
		\caption{The Overall architecture of the proposed model for one transaction}
		\label{fig:arch}
	\end{figure}
	In this section, we delineate our proposed model. We begin with an overview of the overall architecture (Figure~\ref{fig:arch}), then describe each component’s algorithms and formulations in detail. Prior to initiating the steps of the model, it is imperative to initialize the blockchain consensus committee.
	
	The consensus algorithm in blockchain ensures that network participants agree on the state of the blockchain and its transactions~\cite{37}. Before recording a transaction in a block, nodes undergo a validation process called consensus~\cite{sciot2024cons}. Various consensus algorithms exist, including committee-based approaches where record-keeping rights are entrusted to an elected committee of block producers.
	
	In this research, we propose a committee-based consensus algorithm. In this approach, a committee of $K$ nodes collectively evaluates proposals for inclusion in the blockchain. Committee members aggregate proposals, compute quality scores, and update node reputations and delays. Initially, since nodes lack histories, their quality scores and delays are equal. We randomly select $K$ nodes to form the initial committee.
	
	\begin{figure}[!t]
		\centering
		\includegraphics[scale=0.45]{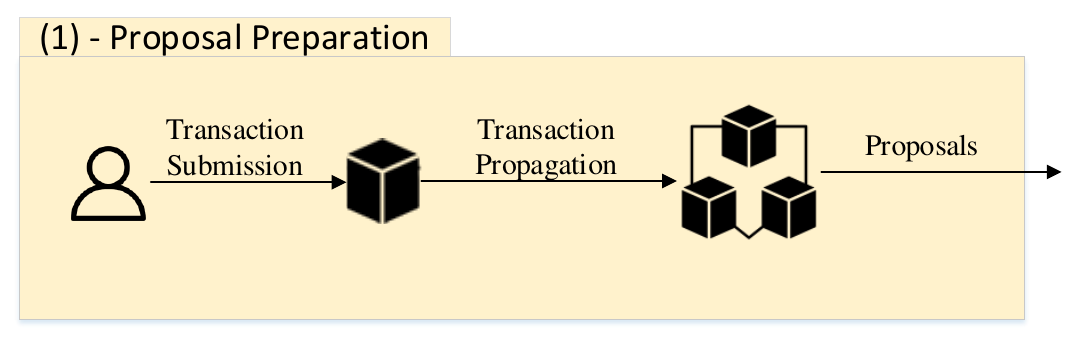}
		\caption{The proposal preparation component of the proposed model}
		\label{fig:prep}
	\end{figure}
	
	\subsection{Proposal Preparation}
	Upon submission of a new transaction, potential changes to node and artifact quality scores arise. This transaction effectively serves as a proposal for altering quality scores. The process involves three sub-steps, as depicted in Figure~\ref{fig:prep}.
	
	\subsubsection{Transaction Submission and Propagation}
	The model begins with generating and submitting a new transaction, indicating the creation of a new artifact in the system (e.g., a published scientific article). This artifact may endorse other artifacts, thereby altering their quality scores. The transaction is submitted to a blockchain node, which verifies its authenticity and validity to prevent spam and fake reports. Once validated, the node assigns a proposal ID, comprising the node ID and a timestamp, and propagates it to all other blockchain nodes.
	
	\subsubsection{Preparing Proposals - Computing quality Scores}\label{sec:preparing-proposals}
	Each node maintains a local copy of the artifact DAG. Upon receiving a new artifact, the node adds it to its DAG, establishes endorsement links, and initializes the artifact's quality score to $S_i^{(0)}$. It then runs a semi-iterative algorithm to update the quality scores of endorsed artifacts, whether directly or indirectly linked. For an artifact, $a_i$, the quality score is computed as follows:

	\begin{align}\label{eq:q}
		S_i &= (1-d)+d \times \big(\sum_{j: j \rightarrow i}\frac{S_j}{L_j}  \big)
	\end{align}
	
	In Equation~\ref{eq:q}, $d$ is a damping factor, typically between 0 and 1, determining the percentage of the score transferred to another artifact.
	When a new artifact endorses others, it can alter their quality scores as per Equation~\ref{eq:q}. We recompute the quality scores of the endorsed nodes, $\{a_j: i \rightarrow j\}$. If these scores change, the update propagates to the next level, triggering a cascade through the DAG that updates the quality scores of all nodes directly or indirectly endorsed by the new artifact. Algorithm~\ref{alg:qpg} details this update propagation process.
	
	\begin{algorithm}
		\caption{Update Propagation Algorithm}\label{alg:qpg}
		\begin{flushleft}
			{
				\textbf{Input: } \\
				\hspace*{0.3cm} a: the new artifact \\
				\hspace*{0.3cm} DAG: the graph of all artifacts and their scores\\
				\hspace*{0.3cm} $\epsilon$: threshold of update propagation\\
				\textbf{Result: } \\
				\hspace*{0.3cm} U: updated nodes\\
				\textbf{Initialization: } \\
				\hspace*{0.3cm} Q : an empty queue\\
				\hspace*{0.3cm} U : an empty list\\
				\hspace*{0.3cm}     $s_i \leftarrow 1$\\
				\hspace*{0.3cm}     $\delta \leftarrow 1$\\
				\textbf{Body: }
			}
		\end{flushleft}
		\begin{algorithmic}[1]
			\State Q.add(a)
			\While{Q is not empty}
			\State t $\leftarrow$ Q.remove()
			\State $\delta \leftarrow \frac{\delta}{L_t}$
			\State $s_t \leftarrow s_t + \delta$
			\State U.add(t,$s_t$)
			\State \If{$\delta \geq \epsilon$}
			\For{$ child \in \{j | i\rightarrow j$\}}
			\State Q.add($child$)
			\EndFor
			\EndIf		
			\EndWhile
			\State Return U
		\end{algorithmic}
	\end{algorithm}
	
	In this algorithm, the constant $\epsilon$ adapts the algorithm to network delays by controlling the depth of update propagation through the DAG. Smaller $\epsilon$ values reduce the depth, while larger values allow deeper propagation.
	
	Initially, $\epsilon$ is set to a very small value and adjusts during the model's execution. We will later explain how $\epsilon$ is adjusted based on processing latency.
	
	\begin{figure}[!t]
		\centering
		\includegraphics[scale=0.35]{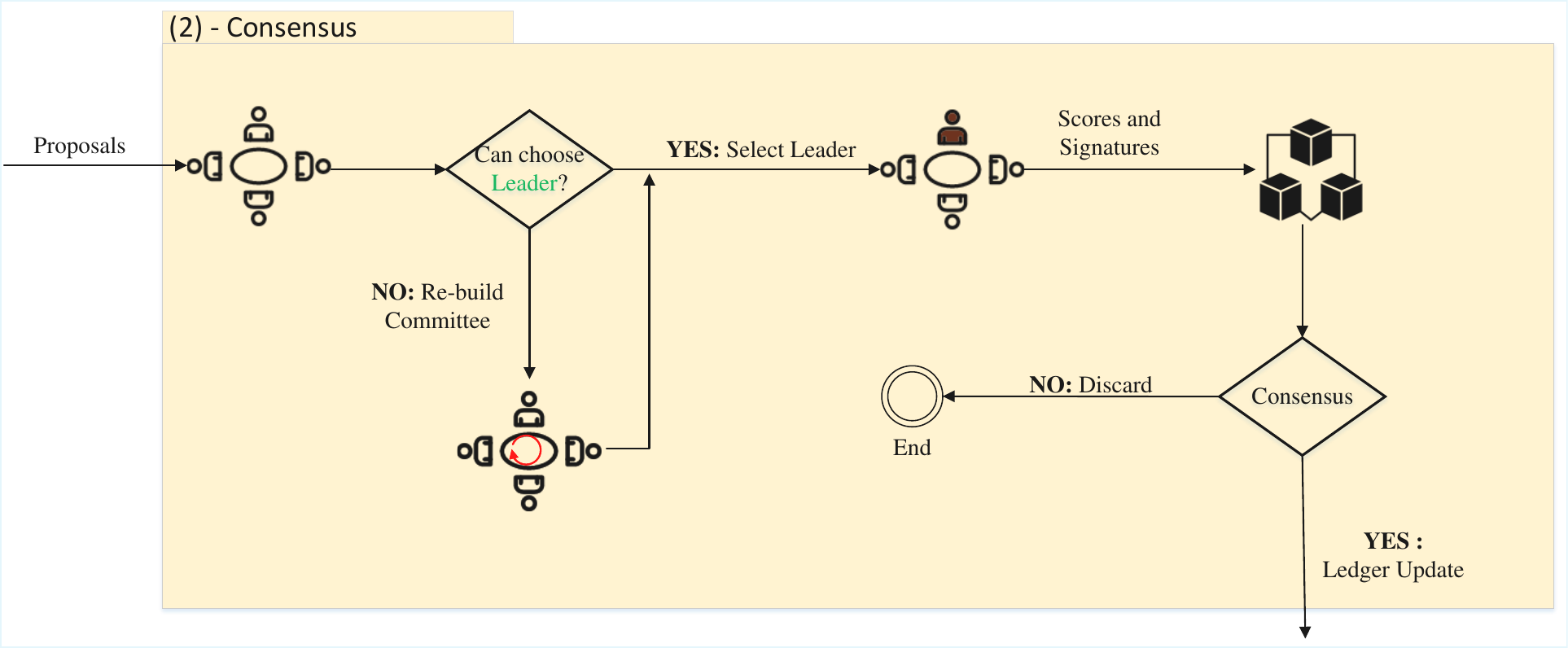}
		\caption{The consensus step of the proposed model}
		\label{fig:cons}
	\end{figure}
	
	After computing the scores, nodes generate a proposal using the proposal ID and propagate it to all committee members.

	\subsection{Consensus}
	In this phase, we collect, validate, and aggregate proposals from blockchain nodes to determine their inclusion on the chain. As shown in Figure~\ref{fig:cons}, the consensus process involves three sub-steps: (i) leader selection and committee reconstruction (if needed), (ii) aggregation and consensus, and (iii) ledger update.
	
	\subsubsection{Leader Selection and Committee rebuild}
	In each round of our consensus algorithm, a committee member is randomly selected as the leader to manage proposal aggregation. Once all members have acted as leader, the committee is rebuilt.
	
	During the rebuild, committee members solve an optimization problem to propose $K$ new nodes. The goal is to maximize the sum of reputation scores while keeping the total delay below a threshold ($\theta$). Each member uses a genetic algorithm to generate a list of proposed nodes, which is then shared with the committee. Nodes proposed by at least 50\% of members form the new committee. If fewer than $K$ nodes are selected, the process repeats until $K$ nodes are chosen. Current committee members are excluded from the new committee.
	
	The genetic algorithm optimizes two parameters: total delay and sum of reputation scores. The delay must remain within limits to ensure performance, while the objective is to maximize reputation scores, enhancing overall efficiency. Algorithm~\ref{alg:cs} details the selection process for new committee members.
	
	\begin{algorithm}
		\caption{Committee selection algorithm}\label{alg:cs}
		\begin{flushleft}
			{
				\textbf{Data: } \\
				C: number of committee members \\
				\textbf{Result: } committee selection 
			}
		\end{flushleft}
		\begin{algorithmic}[1]
			\Require $C > 0$
			\State $Initialization$
			\State $n \gets C$
			\State $N \gets 0$
			\While{$N < C$}
			\For{i $\in$ [1..n]} \Comment{for each node}
			\State $list_i \gets Run  GeneticAlgorithm$
			\State $broadcast(list_i)$
			\EndFor
			\State $committee \gets subscription(list_1..list_n)$
			\State $N \gets NumberOfCommitteeMembers$
			\EndWhile
		\end{algorithmic}
	\end{algorithm}

	\subsubsection{Aggregation and Consensus}
	The aggregation process begins when the committee leader receives proposals from at least $2f+1$ ($f$ is the number of malicious nodes) of the nodes with the same proposal ID. Each committee member, including the leader, then follows these steps to compute the new quality and reputation scores:
	
	\begin{enumerate}
		\item Generate a list of all artifacts that are involved in the proposals of at least $2f+1$ of the nodes.
		\item For each artifact $a$:
		\begin{enumerate}
			\item
			Generate the list of all scores proposed by nodes, denoted by $\Lambda_a$.
			\item
			Compute the mean of $\Lambda_a$, denoted by $\mu_a$.
			\item
			Find the closest member of $\Lambda_a$ to the $\mu_a$, and select it as the quality score of $a$, denoted by $\lambda_a$.
		\end{enumerate}
		
		\item For each node $n$:
		\begin{enumerate}
			\item
			Generate the list of all scores submitted by $n$ in this proposal, denoted as $V_n$. let's use $v_n^a$ to denote the score submitted by $n$ as the quality score of $a$.
			\item
			Compute the $\delta(reputation)$ as follows:
			\begin{align}\label{eq:deltar}
				r &= \sum_{v \in V_n} |\lambda_a - v_n^a|\nonumber\\
				\delta(reputation) & =
				\begin{cases}
					1       		& \text{if } r == 0 \\
					r       		& \text{if } 0 < r \leq 1  \\
					\frac{1}{r}     & \text{if } r > 1
				\end{cases}
			\end{align}
			\item
			The computation of nodes proposal submission delay is calculated due to utilizing in the committee selection process as performance parameter. A higher delay in a node corresponds to diminished performance. Following the submission of proposals by the nodes to the committee, a list of nodes is constructed based on the order of proposal receipt. Committee members disseminate their respective lists within the committee, and the leader constructs a matrix from these lists, wherein each row represents the list of an individual member. Within this matrix, columns are assigned values ranging from $Q$ to 1 (i.e., $Q$,${Q-1}$,…,1) from the first to the last column. The delay of each node is then calculated by following equation:
			\begin{align}\label{eq:proposalDelay}
				Delay_p(n_x) &= \sum_{c \in C_n} Q(n_x)_{c_n}
			\end{align}
			In equation~\ref{eq:proposalDelay} $Delay_p$ is proposal submition delay, $n_x$ is the node x, $c_n$ is column number n and $Q$ is the Q value for $n_x$ at $c_n$.
		\end{enumerate}
	\end{enumerate}
	
	After these computations, the leader sends its results to all committee members for verification. If over 50\% of the committee members approve and sign off the results, leader considers it a consensus; otherwise, the proposal is discarded.
	
	Additionally, the leader monitors process latency. While calculating end-to-end delay in the network is complex, consensus time can serve as a proxy. This time, denoted as $t$, is measured from when a proposal is propagated to when consensus is reached. The leader adjusts the system based on this latency, recomputing $\epsilon$ using the following equation:
	
	\begin{equation}\label{eq:epsilon}
		\epsilon = \frac{T}{1 + e^{-k(t-t_0)}}
	\end{equation}
	In Equation~\ref{eq:epsilon}, $T$ and $t_0$ are application-dependent constants.
	
	After verifying consensus and recalculating $\epsilon$, the leader distributes the new artifact quality scores, nodes' reputation scores, and the updated $\epsilon$ to all blockchain nodes for recording in their local ledgers and DAGs.
	
	\subsection{Ledger Update}
	In this step, nodes receive new scores from the community leader. They first update the artifact quality scores and record the changes in their ledgers as transactions. Next, they update their DAGs based on the latest artifact scores.
	
	Additionally, nodes update and record reputation scores on the chain by incorporating the corresponding $\delta(reputation)$ received from the leader. They also adjust their $\epsilon$ values for subsequent calculations.
	
	\subsection{Summary}
	
	In summary, our model consists of four general steps. (0)~Initializing,  (ii)~submission and propagation of the proposal, (iii)~consensus and aggregation, and (iv)~ledger updates.
	
	\section{Experimentation and Evaluation}\label{sec:eval}
	In this section, we demonstrate the effectiveness of our proposed algorithm in producing scores that are both meaningful and relevant to the context. To the best of our knowledge, the model proposed in this paper is the first of its kind, pioneering a novel approach to quality control in collaborative content generation using blockchain technology. Given the innovative nature of our work, there are no directly related works available for comparison. This unique context necessitates a robust and comprehensive evaluation mechanism to validate our model.
	
	First, we evaluate the performance of our model. We assess its accuracy in computing quality scores, its efficiency in handling large-scale collaborative projects, and its resistance to manipulation attempts. These performance metrics provide a quantitative measure of our model's effectiveness, demonstrating its practical utility in real-world applications. Through this rigorous evaluation process, we aim to establish our model as a reliable and efficient tool for quality control in collaborative content generation.
	
	We then continue with the theoretical validation mechanism and schema comparison with recent studies. This mechanism scrutinizes the meaningfulness of our model by aligning it with the fundamental principles of both blockchain technology and quality control theory. We examine the model's adherence to key blockchain features such as decentralization, immutability, and security, as well as its alignment with quality control factors such as accuracy and resistance to manipulation. This theoretical validation ensures that our model is not only innovative but also grounded in established theories and principles.
	
	\subsection{Experimental Evaluation}\label{sec:experimental-eval}
	In this section, we illustrate the effectiveness of our model in producing scores that are both meaningful and contextually relevant. Our evaluation strategy involves a comparative analysis between our innovative algorithmic framework and two well-established iterative base models: Google PageRank and the HITS model.
	
	Google PageRank and the HITS model are closely related works that have gained significant recognition for their ability to generate meaningful results. They serve as robust benchmarks in our field, providing a solid foundation for comparison. By comparing our algorithm with these renowned models, we aim to highlight the correlation between our results and theirs.
	
	This comparative approach not only underscores the validity of our algorithm but also provides a familiar reference point for understanding its performance. It allows us to demonstrate how our algorithm maintains the strengths of these well-known models while introducing novel features that enhance its applicability and effectiveness. Through this rigorous comparative analysis, we aim to establish the credibility and value of our proposed algorithm in the realm of collaborative content generation.
	
	\subsubsection{Dataset and Experimentation Setup}
	Experimental procedures were executed utilizing the AMiner dataset V11\footnote{https://www.aminer.org/citation}, encompassing reference counts for 1,048,576 scholarly papers. We validate our model using citation data from AMiner, a benchmark in scientometric studies, ensuring our results generalize to real-world academic collaboration ecosystems. From this dataset, we derived distribution parameters: $\mu=19.02$ and $\sigma^2 = 12.17$. Leveraging this distribution, we synthesized a dataset comprising 1,000 artifacts, with reference counts drawn from the distribution of the real-world dataset.
	
	The simulation environment has been developed in the Golang programming language and runs on a computer with the following technical specifications: Processor: Intel-COREi7-8thGen, Memory: 16GB, Operating System: Ubuntu-v22.04
	
	\subsubsection{Rationality of the Quality Scores}
	
	\begin{figure}[!t]
		\centering
		\subfigure[Comparing with HITS]{
			\includegraphics[scale=0.35]{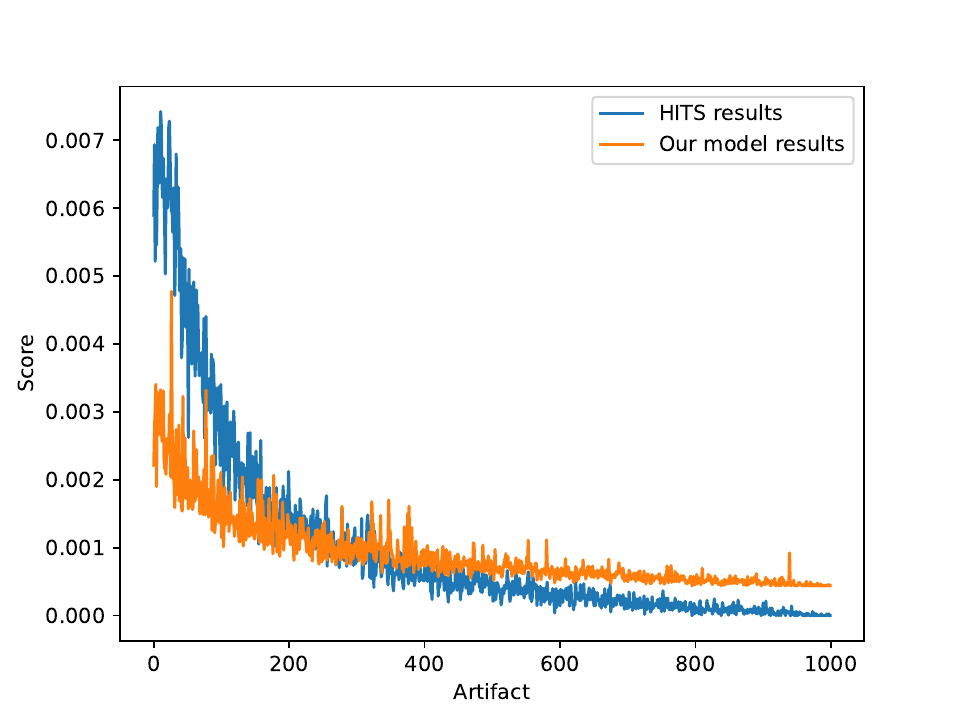}
			\label{fig:cph:a}
		}
		\subfigure[Comparing with PageRank]{
			\includegraphics[scale=0.35]{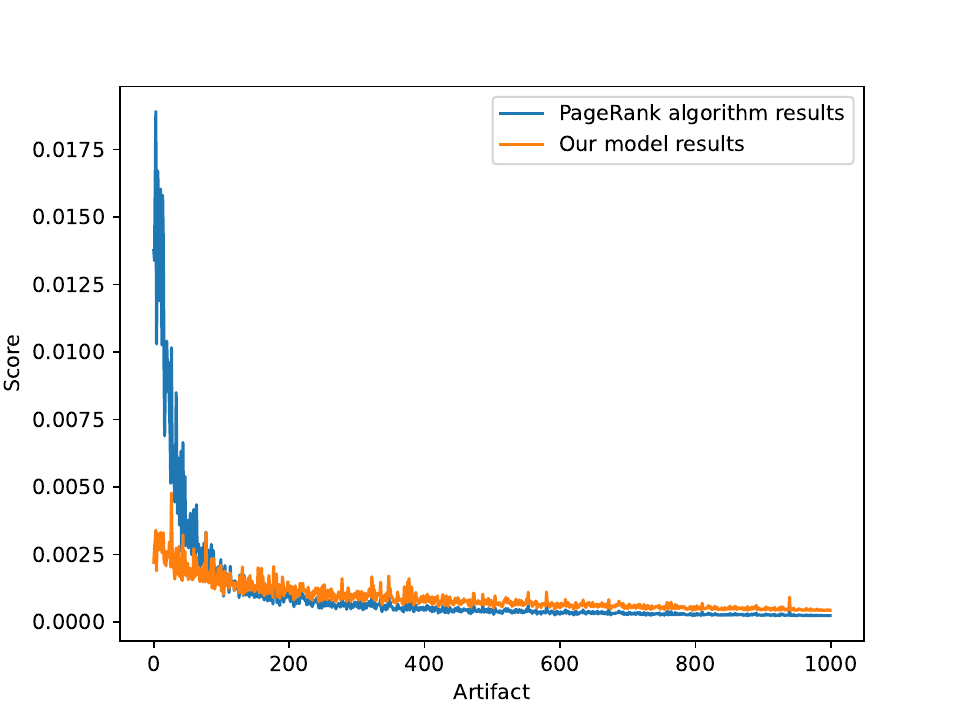}
			\label{fig:cph:b}
		}
		\caption[]{Comparing quality scores (Normalized Citation Impact) computed by proposed model with scores of PageRank and HITS}
		\label{fig:cph}
	\end{figure}

	The comparative analysis results are visually represented in Figure~\ref{fig:cph}, which delineates the correlation between the outcomes produced by our proprietary algorithm and those derived from the established PageRank and HITS models. The horizontal axis denotes the artifact number, while the vertical axis represents the quality scores. Artifacts with smaller numbers are older, whereas those with larger numbers are more recent.
	
	The graphical representation notably exhibits a significant alignment between the results procured from our model and those of the PageRank and HITS models. This convergence accentuates the efficacy of our algorithm in emulating the scoring mechanisms utilized by these well-established models. Quantitatively, our model achieves a 80\% correlation with PageRank results and a 93\% correlation with HITS results, further validating its scoring accuracy.
	
	A distinctive variance between our model and the selected related works lies in the scoring of older artifacts. Both HITS and PageRank assign relatively high scores to older artifacts, predominantly due to indirect endorsements. An indirect endorsement implies that the artifact has not directly cited or endorsed the older artifact, but it may have endorsed artifacts that are distant from the older one, yet it still benefits from these endorsements. We posit that there should be a discernible distinction between direct and indirect endorsements, as well as between close and distant endorsements, and how the artifacts capitalize on these endorsements. The results indicate that our model exhibits a more rational approach from this perspective.
	
	\subsubsection{Throughput and Latency Analysis}
	
	\begin{figure}[!t]
		\centering
		\subfigure[Transaction Confirmation Latency]{
			\includegraphics[scale=0.35]{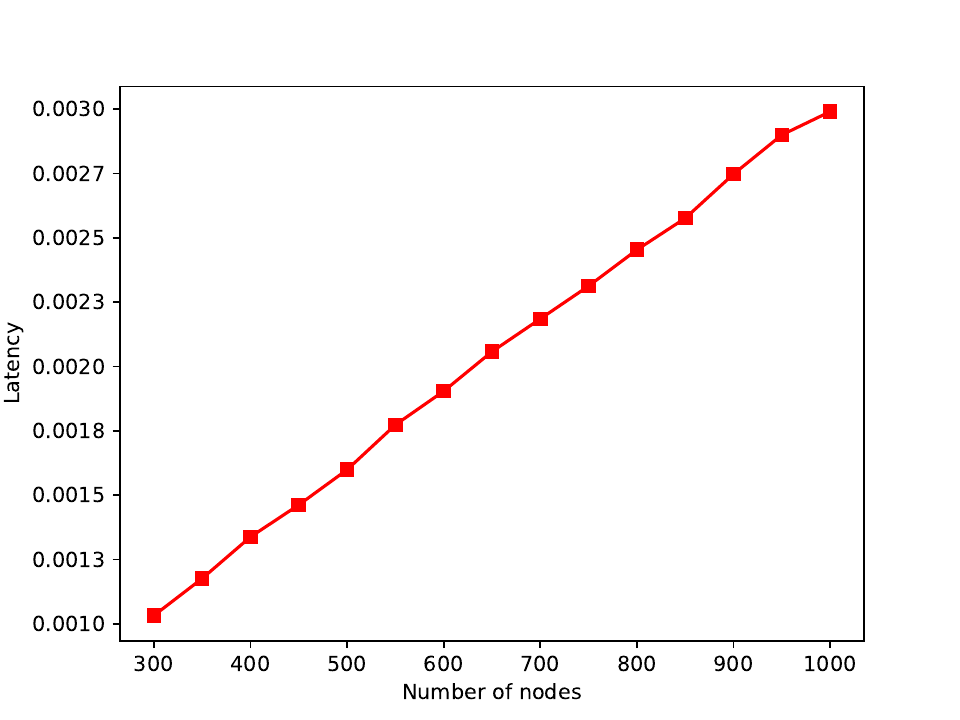}
			\label{fig:tl:a}
		}
		\subfigure[Throughput (Transaction Per Second)]{
			\includegraphics[scale=0.35]{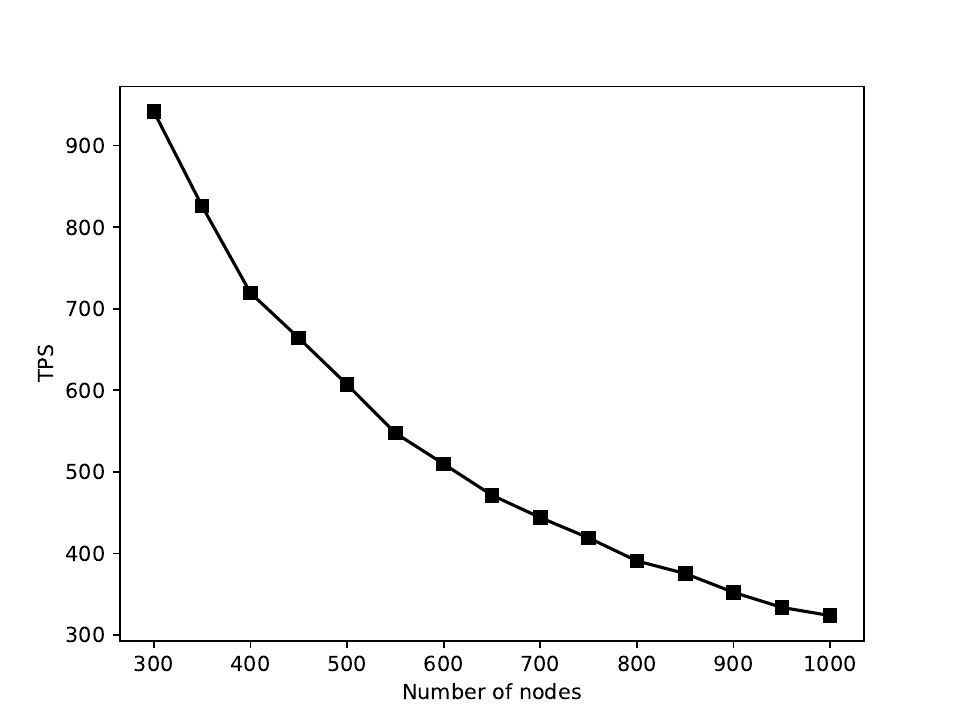}
			\label{fig:tl:b}
		}
		\caption[]{Throughput and transaction confirmation latency of the proposed model}\label{fig:tl}
	\end{figure}
	
	Figure~\ref{fig:tl} provides a detailed representation of the system's throughput and latency respectively. Figure~\ref{fig:tl:a} portrays the system's performance in terms of latency, which is defined as the time interval from the instant a transaction is introduced into the system until it is recorded in the ledger. The graph demonstrates a reasonable latency, even in the presence of a thousand peers, which is considered acceptable.
	
	Our model demonstrates exemplary performance in terms of throughput for the targeted applications. As previously explained, this model is engineered for systems where content is generated collaboratively. Such systems typically do not exhibit a high arrival rate. For example, scientific publishing platforms generally do not have arrival rates that exceed hundreds, nor do they have thousands of peers within their network. As illustrated in Figure~\ref{fig:tl:b}, the proposed model can efficiently process hundreds of transactions even in the presence of a thousand peers, which is an acceptable throughput for the intended application domains.
	
	\subsubsection{Sensitivity to Processing Latency}
	\begin{figure}[!t]
		\centering
		\includegraphics[scale=0.4]{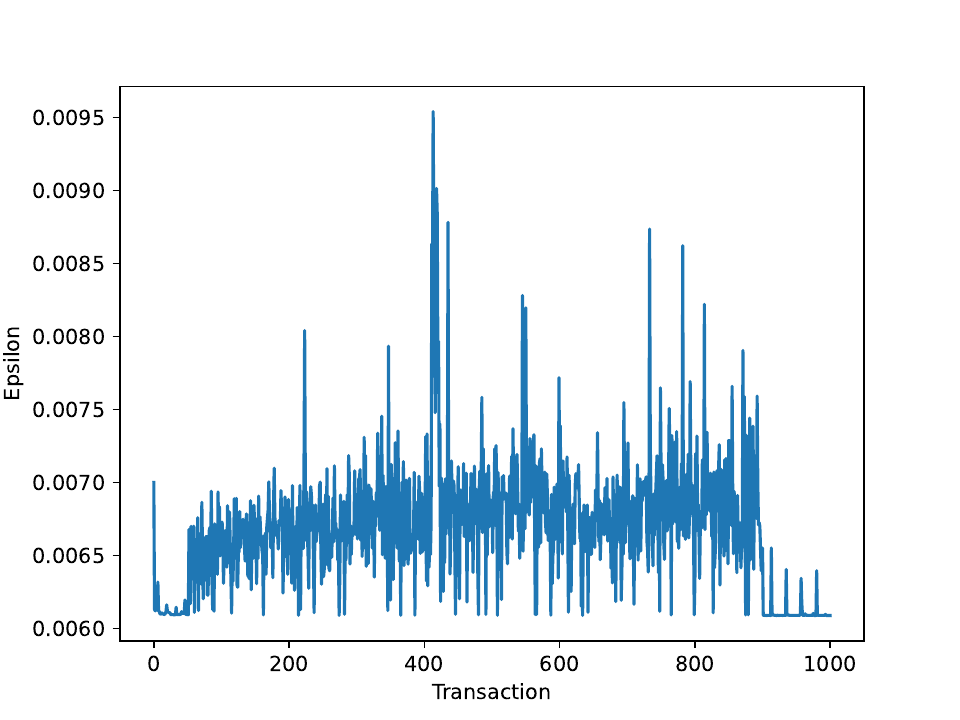}
		\caption{How $\epsilon$ responds to the processing latency}
		\label{fig:epsilon}
	\end{figure}
	
	One of the distinctive features of our system is its sensitivity to processing latency, which is regulated by $\epsilon$ (refer to Equation~\ref{eq:epsilon}). In our experimental setup, we established values for the constants in the equations as follows: $T= 0.05$, $k=50$, and $t_0 = 0.04$  to ensure satisfactory performance. In accordance with these settings, Figure~\ref{fig:epsilon} illustrates how $\epsilon$ responds to latency. The chart reveals that the value of $\epsilon$ undergoes continuous modification during the experimental run, thereby enabling the model to adapt to the latency and maintain an acceptable throughput.
	
	\subsubsection{Security and Robustness}
	In this section, we validate the security and robustness of our model in the face of malicious nodes. At the data recording level, our model utilizes a standard blockchain on each node, in addition to a Directed Acyclic Graph (DAG) for preserving the most recent state of the artifact graphs. These structures inherently exhibit robustness against manipulation, and thus, we do not evaluate them. Instead, our focus is on assessing the robustness of our committee selection algorithm, and we will ascertain its behavior in the presence of malicious members.
	
	Initially, we scrutinize the performance of our approach within a hostile environment that includes malicious nodes in the network. We considered a certain number of malicious nodes within the network and evaluated their impact on the system's performance. We first operate our model in the absence of malicious nodes, and document the results. Subsequently, we run the model with varying percentages of malicious nodes, ranging from 5\% to 65\%, and record their results. Finally, we measure the correlation between the results obtained without malicious nodes and those obtained in their presence. This indicates our approach can tolerate nearly 50\% of the nodes being malicious, which is on par with many Byzantine fault tolerant systems that require a majority of honest nodes.
	
	\begin{figure}[!t]
		\centering
		\subfigure[Impact on the quality scores]{
			\includegraphics[scale=0.35]{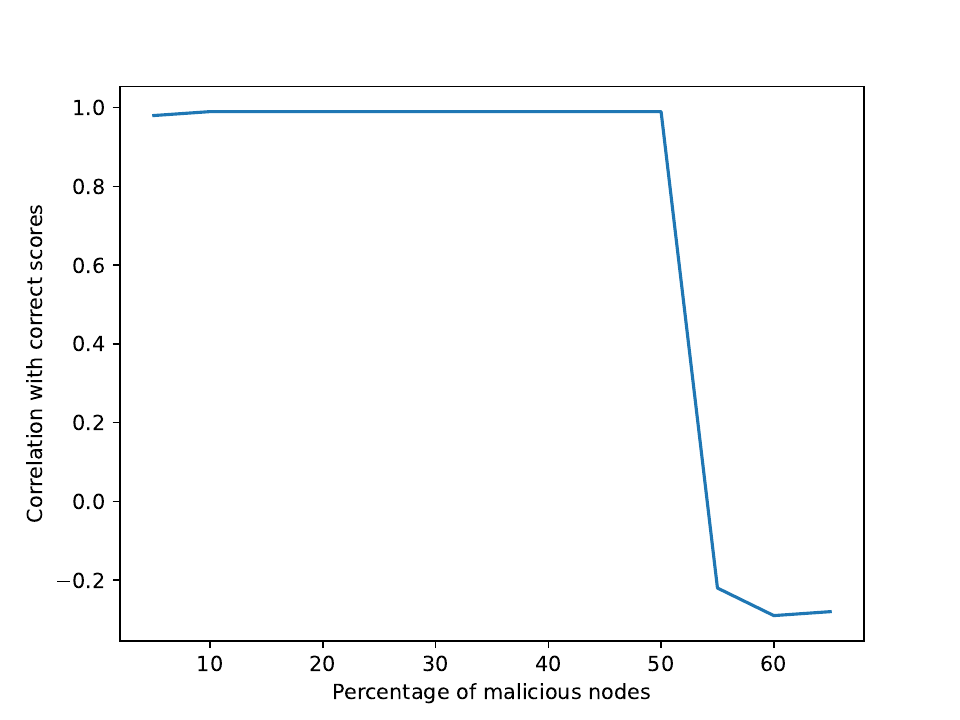}
			\label{fig:ar}
		}
		\subfigure[Impact on the reputation scores]{
			\includegraphics[scale=0.35]{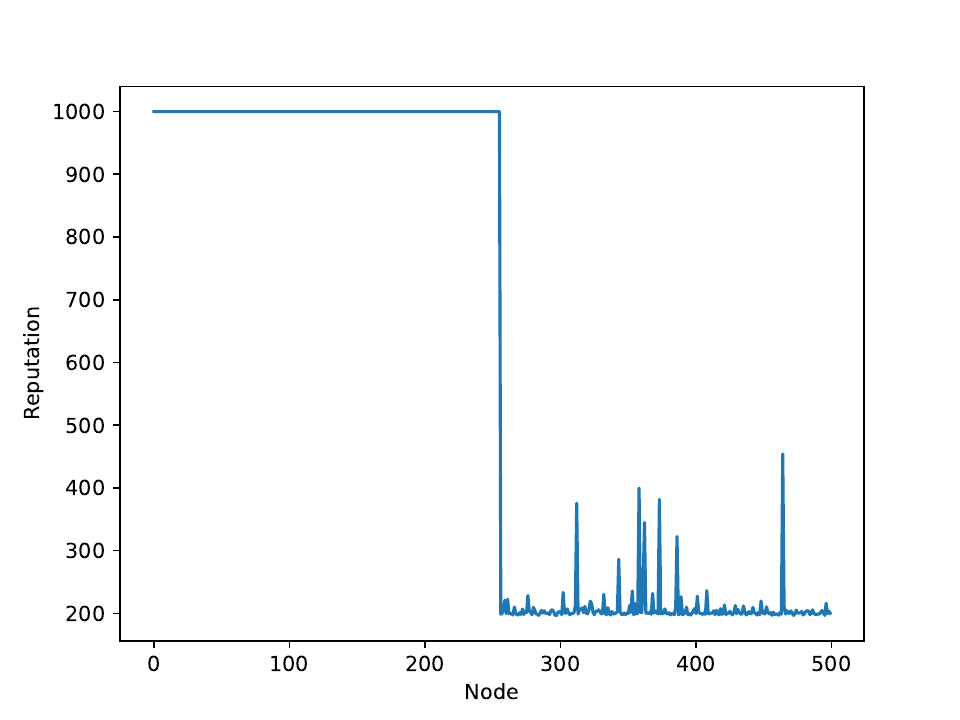}
			\label{fig:r}
		}
		\caption[]{Impact of malicious nodes}\label{fig:rar}
	\end{figure}
	
	It has been observed that with an escalation in the number of malicious nodes from 5\% to 49\%, the quality of the results remains proportionally close to 100 in terms of accurate performance and correlation with the true scores. As depicted in Figure~\ref{fig:ar}, even in the presence of malicious nodes, the system functions correctly, and the scores are computed accurately. Thanks to our consensus and reputation mechanism the presence of malicious nodes does not exert a significant impact on the system's performance. However, with an increase in the number of malicious nodes beyond 49\%, the adverse effect of these nodes on the system's performance becomes more pronounced, and the correlation with true scores diminishes.
	
	Secondly, we scrutinize the performance of our model in terms of its impact on detecting malicious nodes within the network. To accomplish this, we operate our model with 1000 artifacts, and 500 nodes, where the first 250 nodes are honest and the latter half are malicious. Honest nodes exhibit correct behavior and report accurate answers. Conversely, the malicious nodes report random scores for artifacts, as opposed to correct answers.
	
	As illustrated in Figure~\ref{fig:r}, our model accurately distinguishes between honest and malicious nodes. As indicated in Equation~\ref{eq:deltar}, a fully honest node receives $\delta(reputation=1)$ for each transaction. Consequently, the reputation score of honest nodes should be 1000, following the experiments, which is indeed the case. Moreover, the malicious nodes are penalized with significantly low reputation scores, which diminishes their likelihood of entering the committee and manipulating the system.
	We have demonstrated that our model is capable of detecting malicious nodes. In the final part of the evaluation, we illustrate how the committee selection algorithm can identify and prevent malicious nodes from entering the committee.
	
	We assess the robustness of the committee selection algorithm under three scenarios. In the first scenario, 15\% of the network nodes are malicious. This ratio increases to 25\% and 49\% in the second and third scenarios respectively. For a more comprehensive evaluation, we repeated the experiment 1000 times and measured the distribution of malicious nodes in the committee.
	
\begin{figure}[!t]
		\centering
		\subfigure[Scenario 1: 15\% malicious nodes]{
			\includegraphics[scale=0.25]{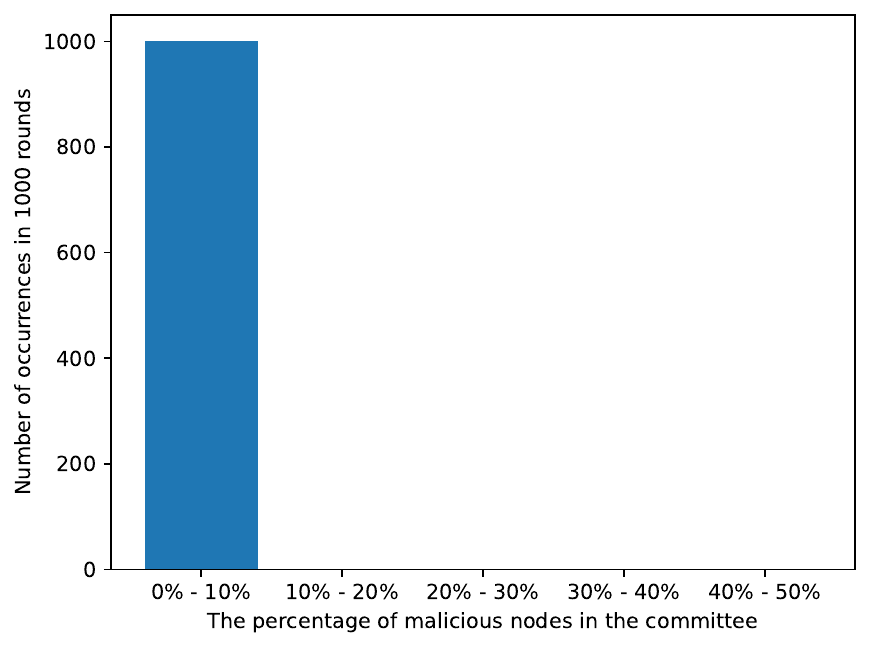}
			\label{fig:15}
		}
		\subfigure[Scenario 2: 25\% malicious nodes]{
			\includegraphics[scale=0.25]{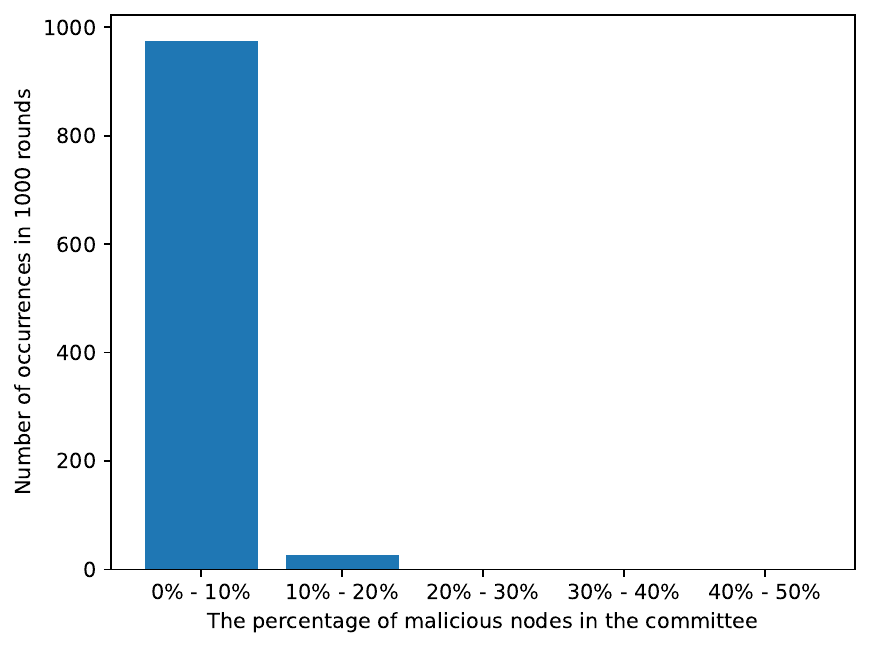}
			\label{fig:25}
		}
		\subfigure[Scenario 3: 49\% malicious nodes]{
			\includegraphics[scale=0.25]{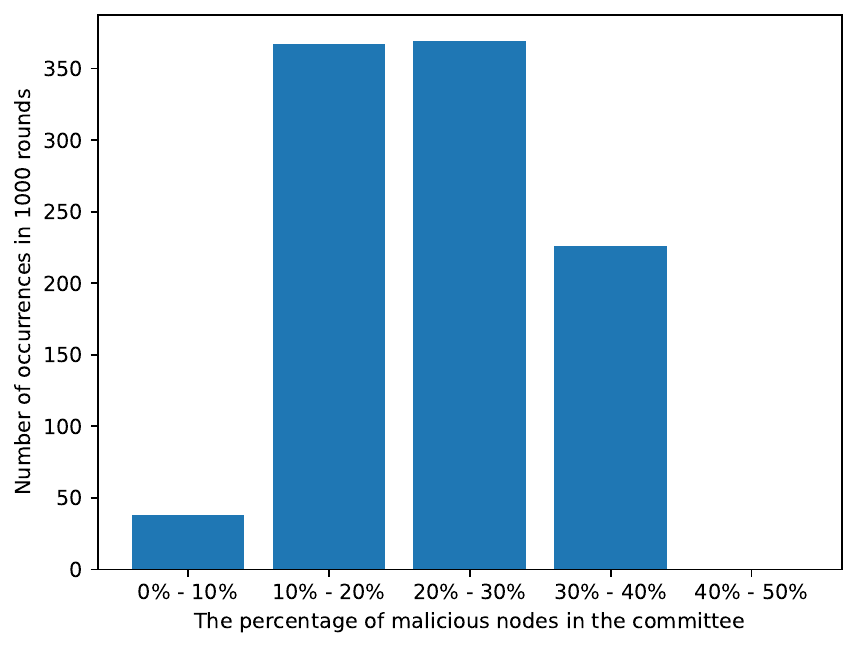}
			\label{fig:49}
		}
		\caption[]{Distribution of malicious nodes in the committee.}
		\label{fig:bd}
	\end{figure}

	Figure~\ref{fig:bd} presents the results of this evaluation. In the charts of this figure, the horizontal axis displays five baskets in the range of 0\% to 50\% that the number of malicious nodes in each run may fall into. The horizontal axis shows the number of times out of 1000, that the number of malicious nodes fall into the specified basket.

	As shown in Figure~\ref{fig:15}, in the first scenario, in all 1000 instances, the number of malicious nodes allowed to enter the committee falls into the [0, 10) basket. In the second scenario, only a few nodes fall into the second basket, which is [10, 20), and the rest are in the first basket. In the third scenario, which is the most severe one, the highest observed percentage of malicious nodes within the committee over 1000 runs is 38
	
	This demonstrates that in all these scenarios, the percentage of the malicious nodes is always less than 40\%, which is easily manageable by our model as shown in Figure~\ref{fig:ar}. This implies that when the number of malicious nodes in the committee increases but does not reach the 50\% threshold, the committee can prevent issues and disruptions and continue to operate in a coordinated and consensus-driven manner. This feature underscores the importance of consensus and cohesive functioning within the committee. When the number of malicious nodes exceeds 50\%, the committee's consensus may be affected, potentially impacting the system's performance. However, until this threshold is reached, the committee can mitigate the negative effects of malicious nodes through consensus and coordination, preserving the quality of performance.
	
	\subsection{Theoretical Validation}
	In this section, we undertake a theoretical validation of our proposed model. Theoretical validation is a critical step in the research process that ensures our model is not only innovative but also grounded in established theories and principles. It involves examining the model's alignment with the underlying theories it is based on, in this case, blockchain technology and quality control theory.
	
	\subsubsection{Single Point of Failure}
	Centralized models, whether they are iterative, heuristic, weighted average, or simple aggregation, inherently have a single point of failure. This is because they rely on a central authority or server to compute and distribute quality scores. If this central authority fails, the entire system can become inoperative. On the other hand, our proposed decentralized model mitigates this risk. By distributing the computation and storage of quality scores across multiple nodes in the blockchain network, we eliminate the single point of failure. Even if one node fails, the system continues to operate seamlessly, ensuring uninterrupted service.
	
	Moreover, the decentralized nature of our model enhances its robustness. In a centralized model, any disruption to the central authority, be it due to technical glitches, cyber-attacks, or other unforeseen circumstances, can compromise the entire system. In contrast, our model's decentralized architecture ensures that the system remains functional and resilient in the face of such disruptions, further reinforcing its superiority over centralized models in terms of avoiding a single point of failure.
	
	\subsubsection{Security and Data Integrity}
	In terms of security and data integrity, centralized models again fall short. They are vulnerable to cyber-attacks, and any breach can lead to the alteration or loss of quality scores. Our decentralized model, however, leverages the inherent security features of blockchain technology to safeguard against such threats. The use of cryptographic hashes in blockchain ensures that once a quality score is recorded, it cannot be altered, thereby preserving data integrity.
	
	Furthermore, the consensus mechanism in our blockchain-based model adds an additional layer of security. It requires the majority of nodes to validate a new quality score before it is added to the blockchain. This makes it extremely difficult for any malicious actor to manipulate the quality scores, as they would need to control the majority of the network, which is practically infeasible. Thus, our model not only ensures data integrity but also offers enhanced security compared to centralized models.
	
	\subsubsection{Transparency}
	Transparency is another area where our decentralized model outperforms centralized ones. In centralized models, the computation of quality scores is often opaque, with contributors having little visibility into how their contributions are evaluated. This lack of transparency can lead to mistrust and discourage participation. Our model, however, ensures complete transparency in the computation and assignment of quality scores. All transactions are recorded on the blockchain and are visible to all participants, fostering trust and encouraging active participation.
	
	Additionally, the transparency of our model extends to its resistance to manipulation. In centralized models, the central authority has the power to alter quality scores, leading to potential bias or manipulation. In contrast, the immutable and transparent nature of blockchain in our model ensures that once a quality score is computed and recorded, it cannot be changed. This transparency and immutability make our model more fair and trustworthy, further highlighting its advantages over centralized models.
	
	\subsubsection{Theoretical Comparison}
	
	To illustrate the novelty and comprehensiveness of our proposed approach, we perform a schema comparison with other recent articles in the Table~\ref{table:comparision}. We have reviewed several influential models and frameworks in the literature and assessed them based on a set of selected parameters. The parameters—Transparency, Decentralization, Dynamic Scoring, Interdependence, Scalability, Immutability, Application Flexibility, Fault Tolerance, Consensus Mechanism, Historical Data Utilization, and Data Privacy—were carefully selected for their crucial roles in enhancing the efficiency and reliability of quality control in online collaboration systems. Transparency fosters trust through verifiable operations; Decentralization removes central control points, strengthening resilience; Dynamic Scoring Allows the system to adaptively assess user performance over time, enabling more responsive decision-making; Interdependence captures the influence of shared interactions in terms of relations between artifacts; Scalability ensures system viability under growth, while Application Flexibility supports diverse use cases. Fault Tolerance safeguards against systemic disruptions; Consensus Mechanisms offer secure agreement protocols; Historical Data Utilization provides insight-driven decision-making; Lastly, Data Privacy protects sensitive user and artifact information, maintaining confidentiality across the network~\cite{37,new9}.
	
	\begin{table*}[!htb]
		\centering
		\caption{Comparison between our model and previous research studies. ($\checkmark$ for addressed and $\times$ for not addressed)}
		\label{table:comparision}
		\resizebox{0.85\textwidth}{!}{%
			\begin{tabular}{ |l|c|c|c|c|c|c|c|c| }
				\hline
				\textbf{} & \textbf{\cite{7}} & \textbf{\cite{AQA}} & \textbf{\cite{31}} & \textbf{\cite{34}} & \textbf{\cite{new2}} & \textbf{\cite{new7}} & \textbf{\cite{new11}} &  \textbf{Our Model} \\
				\hline
				\textbf{Transparency} & \checkmark  & \checkmark & \checkmark & \checkmark & \checkmark & \checkmark & \checkmark & \checkmark \\
				\hline
				\textbf{Decentralization} & $\times$  & $\times$ & $\times$ & \checkmark & \checkmark & \checkmark & \checkmark & \checkmark \\
				\hline
				\textbf{Dynamic Scoring} & \checkmark  & \checkmark & \checkmark & \checkmark & $\times$ & $\times$ & $\times$ & \checkmark \\
				\hline
				\textbf{Interdependence} & \checkmark & \checkmark & \checkmark & $\times$ & $\times$ & $\times$ & \checkmark & \checkmark \\
				\hline
				\textbf{Scalability} & \checkmark & \checkmark  & \checkmark & $\times$ & \checkmark & \checkmark & $\times$ & \checkmark \\
				\hline
				\textbf{Immutability} & $\times$ & $\times$  & $\times$ & \checkmark & \checkmark & \checkmark & \checkmark & \checkmark \\
				\hline
				\textbf{Application Flexibility} & \checkmark & \checkmark  & \checkmark & $\times$ & $\times$ & $\times$ & $\times$ & \checkmark \\
				\hline
				\textbf{Fault Tolerance} & $\times$ & $\times$  & $\times$ & $\times$ & \checkmark & $\times$ & $\times$ & \checkmark \\
				\hline
				\textbf{Consensus Mechanism} & $\times$ & $\times$  & $\times$ & \checkmark & \checkmark & \checkmark & \checkmark & \checkmark \\
				\hline
				\textbf{Historical Data Utilization} & \checkmark & \checkmark  & \checkmark & \checkmark & \checkmark & $\times$ & \checkmark & \checkmark \\
				\hline
				\textbf{Data Privacy} & $\times$ & $\times$  & $\times$ & \checkmark & \checkmark & \checkmark & \checkmark & \checkmark \\
				\hline
			\end{tabular}%
		}
	\end{table*}
	
	As seen in Table~\ref{table:comparision}, our model is the only one (among those compared) that satisfies all listed criteria, whereas others each lack in some aspects. 
	
	\subsection{Discussion: Costs and Delays}
	\subsubsection{Transaction Costs}
	Implementing a blockchain-based system for quality control in collaborative content generation (CCG) systems can incur various transaction costs, particularly when utilizing public blockchains. Public blockchains often require transaction fees to process and validate transactions, which can accumulate over time. However, there are blockchain platforms with minimal fees that could mitigate these costs.
	
	A more efficient approach is to design an exclusive blockchain tailored for the CCG system, specifically a consortium blockchain. Consortium blockchains, unlike public blockchains, do not impose transaction fees, as they are governed by a group of pre-selected nodes and stakeholders. This setup eliminates public transaction fees while preserving the immutability and security benefits of blockchain, since multiple organizations collectively maintain the ledger.
	
	\subsubsection{Time Complexity}
	The time complexity of processing transactions in a blockchain-based system is another critical consideration. As the number of transactions increases, the system may experience delays in processing and validating these transactions. However, in the context of CCG systems, where content generation is a collaborative effort among humans, the rate of content creation is typically manageable within the proposed model.
	
	The semi-iterative algorithm employed in our system is designed to handle the expected transaction volume efficiently. By adapting to network delays and maintaining a predefined accuracy threshold, the system ensures timely and accurate processing of quality scores. This adaptability makes the model robust against potential delays, ensuring that the quality control process remains effective even as the volume of transactions grows.
	
	In summary, while transaction costs and time complexity are important factors to consider, the proposed consortium blockchain model and the adaptive algorithm provide effective solutions to manage these challenges, ensuring a cost-efficient and timely quality control process in CCG systems. Thus, neither transaction fees nor processing delays are expected to impede the practical deployment of our model.
	
	\section{Conclusion}\label{sec:conc}
	
	In this manuscript, we present a novel blockchain-based quality control framework for Collaborative Content Generation (CCG) systems, leveraging a semi-iterative algorithm to compute transparent, robust quality scores for artifacts. By addressing critical informetric challenges such as citation manipulation and centralized metric vulnerabilities, our model ensures decentralized trust and adapts to processing latency, making it suitable for diverse scholarly ecosystems. Comparative evaluations against PageRank and HITS, coupled with robustness tests against malicious nodes, confirm the rationality and reliability of our scores. This work advances informetrics by offering a scalable, tamper-proof alternative to traditional metrics, fostering transparency in collaborative scholarship. Future research will integrate our framework with ORCID and Crossref for decentralized metadata verification and incorporate altmetrics to develop composite impact indicators. We invite the informetrics community to explore this decentralized paradigm to redefine quality assessment in scholarly communication.
	
	\section*{Acknowledgements}
	During the preparation of this work, the author(s) used Grammarly and ChatGPT in order to enhance the readability and clarity of the article through spelling correction and linguistic improvement. After using this tool, the author(s) reviewed and edited the content as needed and take(s) full responsibility for the content of the publication.


	
\end{document}